\documentstyle[aps,prl,epsfig]{revtex}   
\begin{document}
\draft
\twocolumn[\hsize\textwidth\columnwidth\hsize\csname@twocolumnfalse%
\endcsname 
\title{Dzyaloshinski-Moriya Interaction in the 2D Spin Gap System
SrCu$_{2}($BO$_{3}$)$_{2}$}
\author{O. C\'epas$^{1}$, K. Kakurai$^{2,*}$, L.P. Regnault$^{3}$, 
T. Ziman $^{1}$, J.P. Boucher$^{4}$,
N. Aso$^{2}$, M. Nishi$^{2}$, H. Kageyama$^{5}$, Y. Ueda$^{5}$.}
\address{
$^{1}$Institut Laue Langevin, BP 156, F-38042 Grenoble cedex
9, France.\\
$^{2}$ Neutron Scattering Lab., ISSP, University of
Tokyo, 106-1, Shirakata, Tokai, Ibaraki, 319-1106,
Japan. \\
$^{3}$ DRFMC-SPSMS, Laboratoire de Magn\'{e}tisme et de
Diffraction Neutronique, CEA, F-38054 Grenoble cedex 9,
France.\\ 
$^{4}$Laboratoire de Spectrom\'{e}trie Physique,
Universit\'{e} J.  Fourier, BP 87, F-38402 Saint Martin
d'H\`{e}res cedex, France. \\
$^{5}$ Materials Design and Characterization Lab., IRSP, University of Tokyo, 5-1-5 Kashiwa, Chiba, 277-8581, Japan.}

\date{\today}
\maketitle

\begin{abstract}
The Dzyaloshinski-Moriya interaction partially lifts the magnetic
frustration of the spin-1/2 oxide $\rm SrCu_2(BO_3)_2$. It explains
the \textit{fine structure} of the excited triplet state and its
unusual magnetic field dependence, as observed in previous ESR and new
neutron inelastic scattering experiments. We claim that it is mainly
responsible for the dispersion. We propose also a new mechanism for
the observed ESR transitions forbidden by standard selection rules,
that relies on an instantaneous Dzyaloshinski-Moriya interaction
induced by spin-phonon couplings.
\end{abstract}
\pacs{PACS numbers:}
]

Strontium Copper Borate ($\rm SrCu_2(BO_3)_2$) is a new example of a
magnetic oxide with a spin gap \cite{Kageyama99}, with a ground state
well described as simply a product of magnetic dimers in two
dimensions on the bonds giving the strongest magnetic exchange
\cite{Ueda}.  The weaker exchanges are frustrated by the geometry and,
as shown by Shastry and Sutherland\cite{Shastry}, the ground state of
the isotropic Hamiltonian
% that accurately
%describes its thermodynamics \cite{Miyahara} 
is independent of the value of the weaker exchange, up to a
critical value.  The excitations, however, are not
purely local and cannot be explicitly given. Recent
experiments by ESR \cite{Nojiri99} and neutron inelastic scattering
presented here show how in fact there are spin anisotropies needed for
an accurate description of the dynamics. 
We shall show the corrections to the ground state are needed that,
while small, will be necessary to understand many physical properties.
For example at finite external magnetic field $\rm SrCu_2(BO_3)_2$
appears to exhibit a number of finite magnetization plateaux
\cite{Kageyama99,Plateau}, and the anisotropies will determine the
observability of plateaux in different field directions.  Furthermore
$\rm SrCu_2(BO_3)_2$ is believed to be close in parameter space to a
quantum critical point whose nature is somewhat controversial, and
while the anisotropies are small they may be essential to its nature.

\par For spin ${1\over 2}$ the leading anisotropic terms are of form
Dzyaloshinski-Moriya \cite{DM} and exchange anisotropy.  The former is
particularly relevant, since it may not be frustrated even if the
isotropic exchange is.  While a small Dzyaloshinski-Moriya interaction
should not destroy a gap generated by larger isotropic interactions it
modifies the pure locality of the ground state correlations and
delocalizes the first triplet excitation. This is because it appears
in lower order in perturbation theory than the frustrated isotropic
interactions.  In this paper we predict the Dzyaloshinski-Moriya
interactions that should be expected in $\rm SrCu_2(BO_3)_2$ from the
structure, and show that they do indeed explain new features of the
excitations observed with ESR and neutron inelastic scattering
experiments.

\begin{figure}[htbp] 
\centerline{
\psfig{file=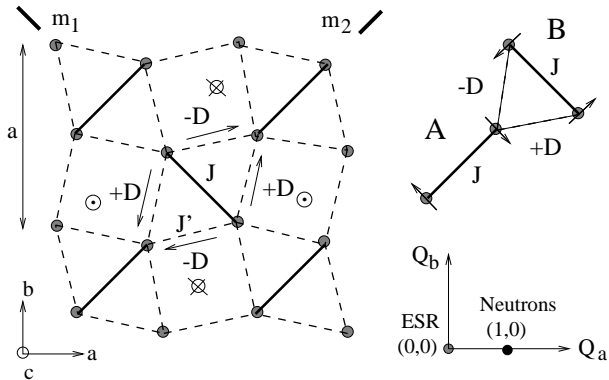,width=8cm}}
\vspace{0.2cm}
\caption{Spin model including the Dzyaloshinski-Moriya interactions whose
vectors are perpendicular to the plane. The arrows show the order of
the spins in the expression $\bf{D}.(\bf{S}_i \times \bf{S}_j)$.  
At right, the unit cell of dimers $A$ and $B$ with the classical
groundstate with $J^{\prime }=0$, and definition of the reciprocal space.}
\label{model}
\end{figure}

\begin{figure}[htbp] 
\centerline{
\psfig{file=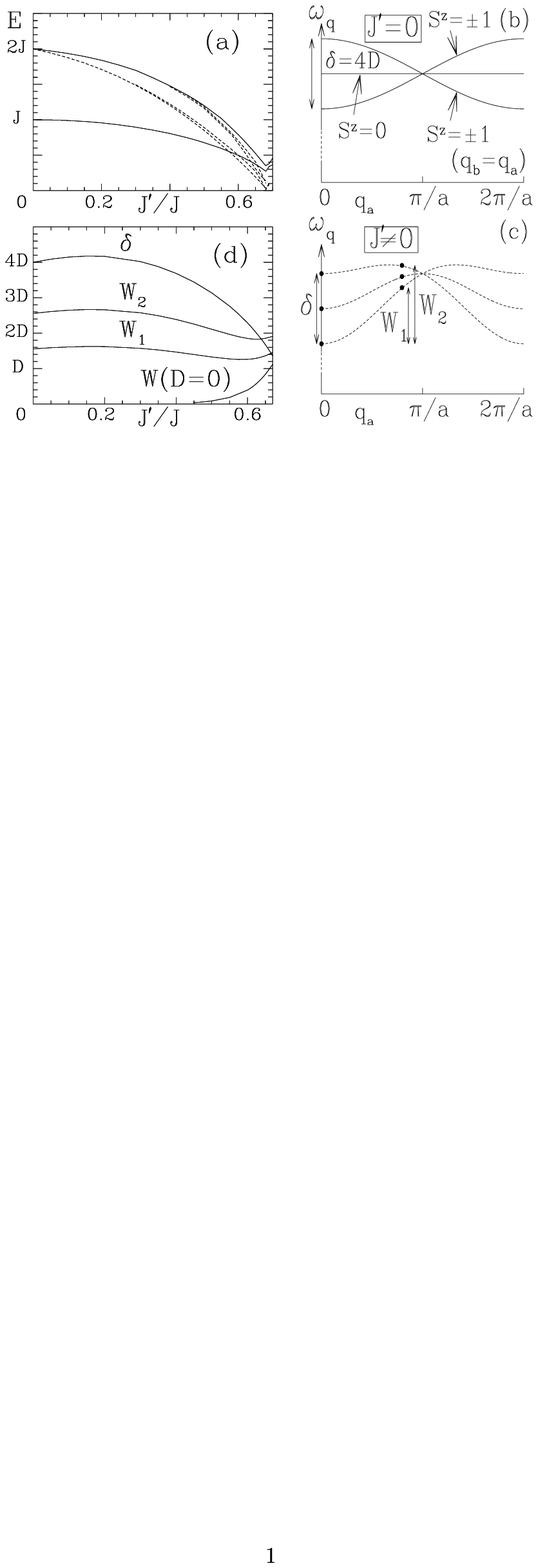,width=8.5cm}
}
\vspace{0.2cm}
\caption{(a) Gaps to the first excited states at ${\bf{q}}=0$ for
$D=0$ (solid line triplets and dashed line singlets), calculated by exact numerical diagonalization. (b) The
effect of the Dzyaloshinski-Moriya interaction on the first triplet for small $J^{\prime}$. 
(c) For finite $J^{\prime}$, the exact diagonalization gives the energies of two reciprocal points (the dots). 
(d)Renormalization of the splitting $\delta$ and the bandwidths $W_1$ and $W_2$.}
\label{resume}
\end{figure}

Miyahara and Ueda\cite{Ueda} have introduced the frustrated
Shastry-Sutherland model

\begin{equation}
H=J\text{ }\sum_{nn}{\bf S}_{i}{\bf S}_{j}+J^{\prime }\text{ }\sum_{nnn}{\bf %
S}_{i}{\bf S}_{j}  \label{1}
\end{equation}
for $\rm SrCu_{2}(BO_{3})_{2}$, with $S=1/2$ and where nn stands for
nearest neighbor spins and nnn for next nearest neighbors. The lattice
is shown in fig. \ref{model}. $J=85K$ and $J^{\prime }=54K$ are
antiferromagnetic interactions estimated from the susceptibility and
the gap \cite{Ueda}. The spectrum of spin excitations has several
interesting features \cite{Kageyama2000}, in particular the existence
of singlet bound states \cite{Lemmens2000}. The figure \ref{resume}a
summarizes the gaps to the first excited states, calculated by exact
numerical diagonalization of the finite size system ($20$ spins). In
addition to the triplet state (solid line) calculated in \cite{Ueda},
the energies of the two lowest singlet states (dashed lines) are given
as a function of $J^{\prime }/J$. These roughly agree with recent
calculation \cite{Knetter}: we find a ratio $J^{\prime}/J\simeq0.62$,
comparing \cite{Georges} the calculated ratio of the energy of the
Raman-active singlet (which is not the lowest singlet \cite{Knetter})
to the triplet excitation with the same ratio determined by experiment
\cite{Kageyama2000,Lemmens2000}.  We find that the energy of the
lowest singlet crosses the ground state at
$(J^{\prime}/J)_c\simeq0.68$ (while the triplet state remains gapped),
in agreement with the transition recently predicted \cite{Koga2000},
but no evidence for the reported $S=1$ instability \cite{Knetter}.
Below this value, the ground state of (\ref{1}) is simply a
product of localized singlets.

\par Anisotropic behavior of the first triplet energy, according to
the direction of an external magnetic field, first appeared in ESR
data \cite{Nojiri99}, and cannot be explained by the fully isotropic
model. We show that the Dzyaloshinski-Moriya coupling, which occurs in
low-symmetry crystals, explains these features.  In $\rm
SrCu_{2}(BO_{3})_{2}$, an almost perfect center of inversion at the
middle of the dimer bonds forbids the Dzyaloshinski-Moriya
interactions between the two spins of a dimer. Each dimer is, however,
separated from the neighboring dimer by a $\rm BO_{3}$ triangle for
which there is no center of inversion at the middle of the bond,
allowing such an interaction between the spins of the nearest neighbor
dimers.  As the copper $({\bf a} {\bf b})$ plane is approximately a
mirror plane for the crystal structure the main components of ${\bf
D}$ must be perpendicular to the plane \cite{DM}, thus lying along the
${\bf c}$ axis. We will neglect the other components, expected to be
smaller. Furthermore, using mirror planes perpendicular to the copper
plane and passing through the dimers ($m_1$ and $m_2$ in the
fig. \ref{model}), we find an alternation of the $\bf{D}$ vector from
bond to bond.  Finally, the mapping of one dimer onto the next one
fixes the whole pattern of $\bf{D}$ vectors (fig. \ref{model}).  We
therefore predict an anisotropic term to the Hamiltonian:

\begin{equation}
H_{DM}=\sum_{nnn} \pm D{\bf e}_{c}.({\bf S}_{i}\times {\bf S}_{j})  
\label{2}
\end{equation}

\noindent
where ${\bf S}_{i}$ and ${\bf S}_{j}$ are spin operators which belong
to next nearest neighbors, the sign $\pm$ depends
on the bond (see fig. \ref{model}). ${\bf e}_{c}$ is the
unitary vector in the $c$ direction. The magnitude can be estimated
from Moriya's argument to be $D \sim \frac{\Delta g_c}{g_c}
J^{\prime}$ where $\Delta g_c \simeq 0.28$ has been measured by ESR
\cite{Nojiri99} and $J^{\prime} \simeq 54K$ \cite{Ueda}.  This gives
roughly $D \sim 0.5 \rm meV$.

%%%%%%%%%%%%%%%%%%%%%%%%%%
%  Fig. 1
%%%%%%%%%%%%%%%%%%%%%%%%%%

The Dzyaloshinski-Moriya interaction is not frustrated because of the
alternation of its vector; classically, in the absence of the exchange
interaction $J^{\prime}$, the spins of the neighboring dimers would be
perpendicular.  Quantum-mechanically, the product of singlets on each
dimer is no longer the exact ground state. To first-order the ground
state has corrections proportional to two triplets $\frac{D}{J}
\prod_{a \neq (b,c)} s_a \left( t^{+1}_bt^{-1}_c - t^{-1}_bt^{+1}_c
\right)$. These corrections are, however, small and we checked by
exact numerical diagonalization that the position of the critical
point separating the two non-magnetic ground states is not appreciably
changed.

In order to calculate the effect of the Dzyaloshinski-Moriya
interaction on the first triplet excitation, we start from small
$J^{\prime}$ and from the exact ground state for $\bf{D}=0$.  At
$J^{\prime}=D=0$, the first excited state is a localized triplet which
is separated from the ground state by an energy gap $J$.  Because of
the frustration, a finite dispersion should appear only at the sixth
order in $J^{\prime}/J$ \cite{Ueda}.  In contrast, the
Dzyaloshinski-Moriya interaction is not frustrated and the degeneracy
of the localized triplets is lifted to first order in $D$. We consider
the two inequivalent triplet states for the dimers $A$ and $B$
(fig. 1). To linear order in $J^{\prime }$ and $D$, we can write the
hamiltonian for the sector of total spin $S^{z}=+1$ (resp. $-1$) as a
$ 2\times 2$ matrix projected on the basis vectors: (i) All dimers are
in singlets bar one, which is in the triplet state, spin $+1$
(resp.$-1$). This triplet is on the sublattice of dimers $A$. (ii) The
same with the triplet on the sublattice of dimers $B$. One obtains:

\begin{equation}
{\cal H}^{S^z=\pm 1} =  
\left(
\begin{array}{cc} 
J 			& \mp 2iD f({\bf{q}}) \\ 
\pm 2iDf({\bf{q}}) 	& J 
\end{array}
\right)
\label{matrix}
\end{equation}

\noindent
where $f({\bf{q}})= \cos(q_a a/2) \cos(q_b a/2) $. The dispersion of
the two modes $\pm$ (each is twice degenerate with $S^z=\pm 1$) is
therefore proportional to $D$:
$\omega_{\bf{q}}^{\pm}(S^z=+1)=\omega_{\bf{q}}^{\pm}(S^z=-1) = J \pm
2D \cos(q_a a/2) \cos(q_b a/2) $.  The Dzyaloshinski-Moriya
interaction has no effect on the $S^z=0$ component of the triplet, so
that its energy remains equal to $J$ [$\omega_{{\bf{q}}}(S^z=0)=J$]
(fig. \ref{resume}b).

%%%%%%%%%%%%%%%%%%%
% Fig. 2
%%%%%%%%%%%%%%%%%%

\begin{figure}
\centerline{
\psfig{figure=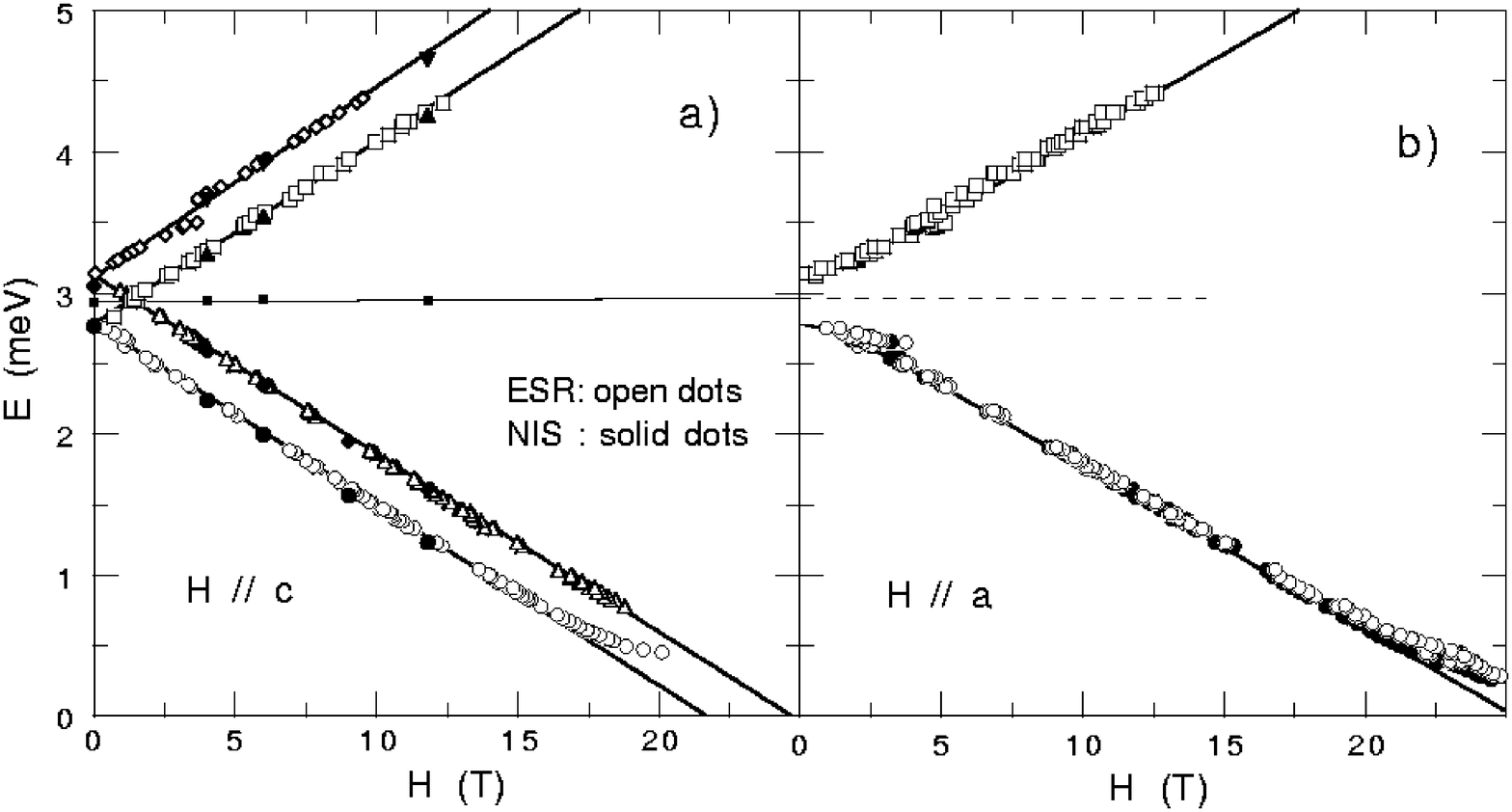,width=8.5cm}
}
%\vspace*{-10.5cm}
\caption{The magnetic field dependence of the triplet energies: (a)
$\bf{H} \parallel c$, (b) $\bf{H} \parallel a$ ; ESR data from
\protect\cite{Nojiri99}, the dashed line comes from
\protect\cite{Room}.  The theory of Dzyaloshinski-Moriya interaction
explains these results (solid lines).}
\label{ESRNeutrons}
\end{figure}

In particular, at ${\bf{q}}=0$, we have two upper (resp. lower) modes
$S^z=\pm 1$ with $\omega(S^z=\pm 1)=J+2D$ [$\omega(S^z= \pm
1)=J-2D$]. A magnetic field parallel to $z \parallel \bf{D} \parallel
c$ splits these modes in four branches. This is in agreement with ESR
when the magnetic field is parallel to the $c$-axis
(fig. \ref{ESRNeutrons}a).  In such an experiment, however, an $S^z=0$
state can not be seen as the external magnetic field is tuned to
adjust the energy of the state to the frequency of the propagating
wave. 
%A state whose energy does not vary with the field will not
%appear in the spectrum.  
We have, however, a clear prediction for the
energy of the mode $S^z=0$: it remains exactly at the middle of the
four $\omega(S^z=0)=J$.  Such a crucial test can be made by neutron
inelastic scattering experiments: we come back to this point next.

A transverse magnetic field leads to the diagonalization of a $6
\times 6$ matrix which reduces to two equal $3 \times 3$ matrices in
the basis of the zero-field eigenstates. Therefore, the energies
remain twice degenerate in a transverse magnetic field and are given
by: $ \omega_{{\bf{q}}}^{\pm} = J \pm
\sqrt{4D^2f^2({\bf{q}})+(g_{\perp} \mu_B H_{\perp})^2}$,
$\omega_{{\bf{q}}}^0 = J$.  This form also fits the ESR results in
transverse magnetic field (fig. \ref{ESRNeutrons}b), apart a small
splitting in the high field regime which may be accounted by the
differences in the $g$ tensors from the dimer $A$ and $B$.

%%%%%%%%%%%%%%%%%%%%%%%%%%%%
% Fig. 3
%%%%%%%%%%%%%%%%%%%%%%%%%%%%

These expressions are first-order in $J^{\prime}/J$ (although
corrections like $DJ^{\prime}/J^2$ may appear), which is not accurate
for $\rm SrCu_2(BO_3)_2$ ($J^{\prime}/J \simeq 0.62$).  We performed
exact numerical diagonalization to see how these results survive in
the strong coupling limit.  We know for instance that the gap between
the ground state and the first triplet state is renormalized to order
$(J^{\prime}/J)^2$ \cite{Ueda}. In addition, we find, for $D=0$, a
finite bandwidth $W$ for the triplet dispersion. This bandwidth is
smaller than that found previously from a perturbative treatment
\cite{Oitma}. We interpret this difference as breakdown of the
perturbative series close to the quantum critical point.  When $D \neq
0$, we calculated again the splitting $\delta$ between lower and upper
modes at ${\bf{q}}=0$. The result $\delta(J^{\prime}/J \rightarrow
0)=4D$ is renormalized by finite $J^{\prime}$ (fig.
\ref{resume}d). Taking the relevant parameters, we find
$\delta(J^{\prime}/J=0.62) \simeq 2.0D$.  On the other hand, the
dispersion is changed from that calculated in the limit of small
$J^{\prime}$: the energy of the $S^z=0$ mode acquires a small
dispersion $W(D=0)$ (fig. \ref{resume}c,d), though smaller than the
$S^z=\pm 1$ dispersions.

\begin{figure}
%\vspace*{-2cm} 
\centerline{
\psfig{figure=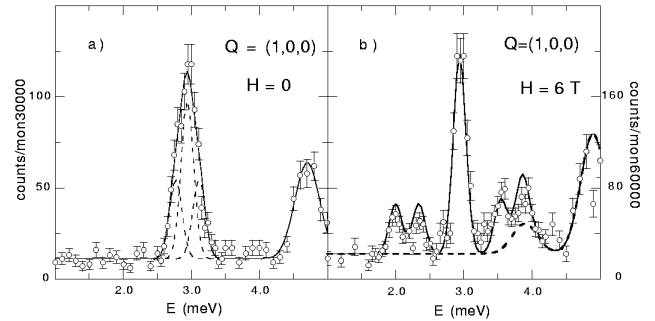,width=8.5cm}
}
%\vspace*{-8.5cm} 
\caption{Scans in the neutron experiment at ${\bf
Q}=(1,0,0)$: (a) the broad signal comes from three superimposed modes
(dashed lines), (b) Zeeman splitting of the modes.}
\label{scans}
\end{figure}

In order to test the prediction of the energy of the $S^z=0$ mode, we
compare to neutron inelastic scattering measurements in the presence
of a magnetic field, up to $H=12$ T. They have been realized on the
three-axes spectrometer CRG/CEA-IN12 installed on a cold neutron guide
at the ILL.  The final wave vector was kept fixed at $k_{f}=1.55
A^{-1}$. Vertical curved pyrolytic graphite PG(002) monochromator and
horizontally curved PG(002) analyser were used and a cooled beryllium
filter was placed after the sample to eliminate higher-order
contamination. The collimations were $ 40^{\prime }$-open-open. A
single crystal of about $15\times 6\times 6$ mm$^{3}$ was installed in
a cryomagnet with the ${\bf c}$ axis aligned along the vertical
field. In the following, we refer to the $({\bf a}^{*}{\bf b} ^{*})$
reciprocal plane defined in \cite{Kageyama2000} (fig. \ref{model}). In
reciprocal lattice units, the wave vector components are defined as
$Q_{a,b}=q_{a,b}a/(2\pi )$ where $a$ is the lattice parameter.
Examples of energy scans obtained at ${\bf Q}=(1,0,0)$ for $H=0$ and
$H=6T$ are shown in fig. \ref{scans}. In fig. \ref{scans}a, we observe
a broad single peak at $E \simeq 3 \rm meV$, corresponding to the
lowest triplet excitation branch (the signal above $\simeq 4.4 \rm
meV$ corresponds to a higher triplet \cite{Kageyama2000} as well as
the dashed line in fig. \ref{scans}b). For $H=0$, it is of note that
the $3 \rm meV$ mode is appreciably broader than the energy resolution
($\simeq 0.2 \rm meV$ at full-width half-maximum). In
fig. \ref{scans}b, the application of $H$ shows a Zeeman splitting of
this mode in five distinct lines. The five peaks become resolution
limited. Their energies are shown in fig. \ref{ESRNeutrons}a.

%%%%%%%%%%%%%%%%
% Fig. 4
%%%%%%%%%%%%%%%%

The energies of the $S^{z}=\pm 1$ excitations at ${\bf Q}=(1,0,0)$
(solid dots) agree with the ESR field-dependent branches (${\bf
Q}=(0,0,0)$), confirming the translational symmetry of the model.  In
addition, we find the energy of the $S^{z}=0$ mode of the triplet
(solid squares) which can not be seen by ESR. It lies exactly at the
middle of the four other branches, as predicted above.  Furthermore,
the observed intensities of the $S^z=\pm 1$ peaks are all equal and
roughly 1/4 of that of the $S^z=0$ peak (fig. \ref{scans}b).  We
calculate the spin correlation function transverse to the scattering
wave vector which gives the intensities.  Because of the purely
imaginary off-diagonal elements in the matrix (\ref{matrix}), the
upper ($S^{z}=\pm 1$) and lower ($S^{z}=\pm 1$) have the same
intensity at zero field.  Both are split in longitudinal field into
two modes of half the zero-field intensity. Similarly, for the
$S^{z}=0$ sector, there are, in fact, two degenerate modes at zero
field which remain degenerate at finite field with the original
weight. Each weight is twice that of $S^{z}=+1$ since only spin
correlation functions transverse to the momentum transfer contribute
to the intensities. The intensity factor $1/4$ then follows. The
agreement with experiment suggests that the ratio of intensities is a
good test of a mechanism based on couplings of Dzyaloshinski-Moriya
symmetry, even in the regime of large $J^{\prime }/J$.

Since the spectra are well fitted for both directions, we can extract
the coupling $D$ from the zero-field splitting $\delta=0.352 \pm 0.008
\rm meV$. Using the numerical result $\delta \simeq 2.0D$ (fig. 2d),
this gives $D \simeq 0.18 \rm meV$, consistent with the approximate
value from Moriya's formula, as given earlier. We obtain the gap
(which is no longer simply $J$ but the function shown in
fig. \ref{resume}a) and the $g$-factors in both directions:
$g_{\parallel}=2.24 \pm 0.09$ and $g_{\perp}=2.00 \pm 0.09$.

We then discuss the origin of the finite intensities in ESR
\cite{Nojiri99} and infra-red absorption \cite{Room}.  No magnetic
transition can be induced from a purely singlet ground
state. Spin-orbit couplings generate anisotropies that may permit
transitions from the ground state to the original triplet excited
states.  First, note that Dzyaloshinski-Moriya interactions are not
sufficient to explain the observed transitions. Indeed, the magnetic
operator $H_M=h\sum_i S_i^+$ is odd under the symmetry
transformation: the reflection using the mirror plane $m_1$ and the
rotation of $\pi$ around the dimer bond. The ground state and the
triplet state are, however, even under this transformation. The matrix
element therefore vanishes, leading to zero intensity.  Second, the
exchange anisotropy only (both in $J$ and $J^{\prime}$) leaves the
ground state as the product of singlets, leading again to zero
intensity.  With both Dzyaloshinski-Moriya and exchange anisotropy,
the transition can exist but should be very small, of the order of
$\lambda ^{6}$, where $\lambda $ is the spin-orbit coupling.  Note
also that an exchange anisotropy of order $D^{2}/J$ always occurs with
a Dzyaloshinski-Moriya coupling\cite{DM}, but unlike the unfrustrated
case\cite{Aharony}, here it does not eliminate the splitting which is
linear in $D$. Third, we have mentioned that the anisotropy of the $g$
tensors may be responsible for the very small splitting of the modes
in high field. A staggered field between dimers (due to perpendicular
orientations of oxygen ions surrounding copper ions) is written:
$H_{M}= h \Delta g \left(\sum_{i \in A} S_i^{+} - \sum_{i \in B}
S_i^+ \right)$.  This term is odd in the
transformation introduced above and transitions are thus forbidden for
the same reasons.  An intra-dimer staggered field should be very small
giving an intensity we estimate as $\lambda^2 \alpha ^{2}$ where
$\alpha $ - a few degrees - is the tilting angle of the
crystal structure\cite{Smith}.

Consider now \textit{electric dipole transitions} between magnetic
states.  We restrict here to a phonon-assisted mechanism. A purely
electronic mechanism may in fact also apply \cite{Abragam}. Treating
the spin-phonon interactions perturbatively leads to an effective
matrix element between pure magnetic states $ \mid \langle f \mid H_E
\mid i \rangle |^2 $, where the effective operator $H_E$ has now a
correction of spin-orbit origin \cite{moi}: $H_E = \sum_{nn} \gamma
\bf{S}_i . \bf{S}_{j} + {\bf{\eta}}.(\bf{S}_i \times \bf{S}_{j})$.
The first term comes from a transition via a virtual phonon as an
intermediate state. The second term, which comes from the spin-orbit
coupling should be $\mid \eta \mid \sim \lambda \gamma$, with a
direction that depends on details of the phonons.  This mechanism
gives an intensity proportional to $\eta^2$, suggesting that the
observed transitions may be electric dipole. Further polarized
experiments can test this: if the intensity of the transition changes
in rotating the crystal around the magnetic (resp. electric) field of
the wave, the transition would be demonstrated explicitly to be
electric dipole (resp. magnetic).

\par We have shown that the \textit{static} Dzyaloshinski-Moriya
interaction explains the magnetic field behavior of the first triplet
excitation (fig.\ref{ESRNeutrons}) and gives a
dispersion (fig.\ref{resume}), because it is not frustrated.  We
extracted its value from the zero-field splitting, taking into account
a finite $J^{\prime}$: $D=0.18 \rm meV$.  Since the direction
of the main anisotropy is given by $\bf{D} \parallel c$, we
expect clear magnetization plateaux only when the magnetic field is
applied along the $c$-axis. This is consistent with current results
\cite{Plateau}.  The static interaction does not, however, explain the
finite intensity for the ESR transitions. We presented an alternative
mechanism using a \textit{dynamical} Dzyaloshinski-Moriya interaction.

We thank Prof. H.Nojiri for providing us with his ESR data and for
interesting discussions.

%\begin{figure}[htbp]
%\centering
%\parbox{4.5cm}{
%\psfig{file=Fig2a.ps,width=4.5cm,angle=-90} 
%\psfig{file=Fig2d.ps,width=4.5cm,angle=-90}
%}
%\parbox{3.8cm}{
%\psfig{file=Fig2b.ps,width=3.8cm,angle=-90}
%\psfig{file=Fig2c.ps,width=3.8cm,angle=-90} 
%}
%\vspace{0.2cm}
%\caption{(a) Gaps to the first excited states at ${\bf{q}}=0$ for
%$D=0$ (solid line triplets and dashed line singlets), calculated by exact numerical diagonalization. (b) The
%effect of the Dzyaloshinski-Moriya interaction on the first triplet for small $J^{\prime}$. 
%(c) For finite $J^{\prime}$, the exact diagonalization gives the energies of two reciprocal points (the dots). 
%(d)Renormalization of the splitting S and the bandwidths $W_1$ and $W_2$.}
%\label{resume}
%\end{figure}

%\vspace*{2.5cm}

\end{document}